\documentclass[aps,twocolumn]{revtex4}

\usepackage{graphicx}
\usepackage{xcolor}
\usepackage{amsmath}
\usepackage{units}
\usepackage{soul}
\usepackage{supertabular}

\begin{document}

\title{Riemann zeros from a periodically-driven trapped ion}

\author{Ran He$^{1,2}$}

\author{Ming-Zhong Ai$^{1,2}$}

\author{Jin-Ming Cui$^{1,2}$}
\email{jmcui@ustc.edu.cn}

\author{Yun-Feng Huang$^{1,2}$}
\email{hyf@ustc.edu.cn}

\author{Yong-Jian Han$^{1,2}$}

\author{Chuan-Feng Li$^{1,2}$}
\email{cfli@ustc.edu.cn}

\author{Guang-Can Guo$^{1,2}$}

\author{G. Sierra$^{3}$}
\email{german.sierra@uam.es} 

\author{C.E. Creffield$^{4}$}
\email{c.creffield@fis.ucm.es}

\vspace{0.5cm} 

\affiliation{$^{1}$CAS Key Laboratory of Quantum Information, University of Science
and Technology of China, Hefei, 230026, People's Republic of China.}
\affiliation{$^{2}$CAS Center For Excellence in Quantum Information and Quantum Physics, University of Science and Technology of China, Hefei, 230026,
People's Republic of China.}
\affiliation{$^{3}$Instituto de F\'isica Te\'orica, UAM-CSIC, E-28049, Madrid, Spain.}
\affiliation{$^{4}$Departamento de F\'isica de Materiales, Universidad Complutense de Madrid, E-28040, Madrid, Spain.}

\date{\today}

\begin{abstract}
The non-trivial zeros of the Riemann zeta function are central objects
in number theory. In particular, they enable one to reproduce the prime
numbers. They have also attracted the attention of physicists working in
Random Matrix Theory and Quantum Chaos for decades.
Here we present 
an experimental observation of the lowest non-trivial
Riemann zeros by using a trapped ion qubit in a Paul trap,
periodically driven with microwave fields. The waveform of the driving is
engineered such that the dynamics of the ion is frozen when the 
driving parameters coincide with a zero of the real component of the
zeta function. Scanning over the driving amplitude thus enables
the locations of the Riemann zeros to be measured experimentally to a
high degree of accuracy, providing a physical embodiment of these
fascinating mathematical objects in the quantum realm.
\end{abstract}

\maketitle

\section{Main}

The Riemann zeta function $\zeta(s)$ is the Rosseta stone for number theory. The stone, found by Napoleon's troops in Egypt, contains the same text written in three different languages, which enabled the Egyptian hieroglyphics to be deciphered. 
The $\zeta$-function is also expressed in three different ``languages'': 
as the series $\sum_{n} n^{-s}$ 
over the positive integers $n$, as the  product $\prod_p 1/(1 - p^{-s})$  over the prime numbers $p$,  and as the  product  
$\propto \prod_{n}   (1 - s/\rho_n) e^{s/\rho_n}$ over the Riemann zeros $\rho_n$ \cite{edwards}. 
Riemann conjectured in 1859 that these zeros would have a real part equal to a half, 
$\rho_n = \frac{1}{2} + i E_n$,
where $E_n$ is a real number \cite{riemann}. 
This is the famous Riemann Hypothesis (RH), one of the six unsolved Millennium problems, whose solution would amplify our knowledge of the distribution of prime numbers with resulting consequences for  
number theory and factorization schemes \cite{millennium,conrey}. More poetically,
in the words of M. Berry, the proof of the RH would mean that {\em there is music in the prime numbers} \cite{music}. 

One of the most interesting ideas to attack the RH is to show that the $E_n$ 
are the eigenvalues of 
the Hamiltonian of a quantum system. This idea, suggested by P\'olya and Hilbert around 1912 \cite{montgomery73}, began to be taken seriously in the 70s  with Montgomery's observation \cite{montgomery75} that the  Riemann zeros closely satisfy  the statistics of the Gaussian unitary ensemble (GUE). In the 80s Odlyzko \cite{odlyzko} tested this prediction numerically for $10^5$ zeros around the $10^{20}$th zero, 
finding only minor deviations from the GUE. These were explained later by Berry and collaborators \cite{berry86,berry88,BK96} using the theory of quantum chaos, and led him to propose that
the $E_n$ are the eigenvalues of a quantum chaotic Hamiltonian whose classical version contains isolated periodic orbits whose periods are the logarithm of the prime numbers. Much work has been done 
\cite{BK99}-\cite{S19}
to find such a Hamiltonian, but so far without a definitive answer. 

In this Letter we present an experimental observation of the lowest Riemann zeros, 
which is quite different from the spectral realization described above. Our intention
is not to prove the RH, but rather to provide 
a physical embodiment of these mathematical objects by using advanced quantum technology. 
The physical system that we consider is a trapped-ion qubit. 
The ion is subjected to a time-periodic driving field, and consequently its behaviour is described
by Floquet theory, in which the familiar energy eigenvalues of static quantum systems
are generalized to ``quasienergies''. These quasienergies can be regulated by the parameters of the 
driving, in a technique termed {\em Floquet engineering}. In particular,
when the quasienergies are degenerate (or cross) the ion's dynamics is frozen, which can
be observed experimentally.
The Riemann zeta function enters into this construction in the design of the driving field, which 
is engineered to produce
the freezing of the dynamics when the real part of $\zeta(s)/s$, with $s= \frac{1}{2} + i E$, vanishes.
Thus observing the freezing of the qubit's dynamics as the driving
parameters are varied gives a high-precision  experimental measurement of the location of the Riemann zeros.

\subsection{Floquet theory}
We consider a two-level system subjected to a time-periodic driving, described
by the Hamiltonian $H(t) = J \left( \sigma_x + \frac{ f(t)}{2}  \sigma_z \right)$, 
where $\sigma_{x,z}$
are the standard Pauli matrices and $J$ represents the bare tunneling
between the two energy levels. Henceforth we will set $\hbar = 1$, and
measure all energies (times) in units of $ J$ ($J^{-1}$).
As $H(t)$ is time-periodic, $H(t) = H(t + T)$,
where $T$ is the period of the driving,
the system is naturally described within Floquet theory, using a basis of Floquet
{\em modes} and {\em  quasienergies} which can be extracted from the unitary time-evolution
operator for one driving-period $U = \mathcal{T} \mathrm{exp} \left[ -i 
\int_0^T H(t') dt' \right]$ (where $\mathcal{T}$ denotes the time-ordering
operator). 
The Floquet modes, $| \Phi_j (t) \rangle$, are the eigenstates of $U$, and the quasienergies, $\epsilon_j$, are related to the eigenvalues of $U$
via $\lambda_j = \mathrm{exp} \left[ -i T \epsilon_j \right]$.

The Floquet modes provide a complete basis to describe the time-evolution of the system,
and the quasienergies play an analogous role to the energy eigenvalues of a time-independent system.
The state of the qubit can thus be expressed as
$| \psi(t) \rangle = \sum_j \alpha_j \mathrm{exp}\left[-i \epsilon_j t \right] | \Phi_j (t) \rangle$,
where the expansion coefficients $\alpha_j$ are time-independent, and the
Floquet modes are $T$-periodic functions of time.
From this expression, it is clear that if two quasienergies approach degeneracy,
the timescale for tunneling between them will diverge as $1 / \Delta \epsilon$.
Although in general it is difficult to obtain explicit forms for the quasienergies, even for the
case of a two-level system, excellent approximations can be obtained in the
high-frequency limit, when $\Omega = 2 \pi /T$ is the largest energy scale of the problem, that is, $\Omega \gg J$. 
In that case one can derive \cite{cec_prb} an effective {\em static} Hamiltonian,
$H_\mathrm{eff} = J_\mathrm{eff} \sigma_x$, where the effective tunneling is given
by 
\begin{equation}
J_\mathrm{eff} = \frac{J}{T} \int_0^T dt e^{-i F(t)} \ .
\label{effective}
\end{equation}
Here $F(t)$ is the primitive of the driving function, $F(t) = \int_0^t dt' f(t')$, 
and the quasienergies are given by $\epsilon_\pm = \pm \left| J_\mathrm{eff} \right|$.
The eigenvalues thus become degenerate when they are zero, corresponding to
the vanishing of $J_\mathrm{eff}$ and the freezing of the dynamics.
This expression is accurate to first order in $1 / \Omega$, and although in principle higher-order
terms could be calculated using the Magnus expansion, we will work at sufficiently high 
frequencies for this expression to give results of excellent accuracy.

Equation \eqref{effective} is the key to our approach. By altering
the form of the driving, $f(t)$, we are able to manipulate the effective tunneling and 
the quasienergies of the driven system.
Our aim is to obtain a driving function such
that $J_\mathrm{eff}(E)$ is proportional to the real part of $g(E)$ 
with $g(E)= - \zeta \left( 1/2 + i E \right) / \left( 1/2 + i E \right)$,
yielding an effective Hamiltonian whose dynamics is intimately related to the
properties of the $\zeta$-function.
In particular, the effective tunneling will vanish,
an effect termed {\em coherent destruction of tunneling} (CDT) \cite{cdt}, when $E$ coincides with one
of the Riemann zeros. In {\bf Methods} we give the details 
of the mathematical derivation of the driving function, which enables us to obtain a Fourier series for
$f(t)$ (see Fig. \ref{fig:math}d) which can be straightforwardly
programmed into a waveform generator to provide the experimental driving.
We choose to focus on the function $-\zeta(s)/s$ for two fundamental reasons. The
first is that it has a remarkably simple Fourier transform.
This also motivated van der Pol \cite{vdp} and Berry \cite{berry} to use
this function as the basis for physical implementations of the Riemann zeros in diffraction
experiments (in Fourier optics and in antenna radiation patterns respectively). The second reason
is that this function decays slowly as $E$ increases (see   Fig. \ref{fig:math} a ). 
In previous work \cite{pra_1,pra_2} 
we proposed to use Floquet engineering
to simulate the Riemann $\Xi$-function \cite{edwards}. 
Although successful, the extremely rapid decay of the
$\Xi$-function meant that only the lowest two Riemann zeros were resolvable. In contrast,
the slower decay of $-\zeta(s)/s$ should allow many more quasienergy crossings to
be detectable, and thus more zeros to be identified.

\begin{figure*}
\includegraphics[width=1\linewidth]{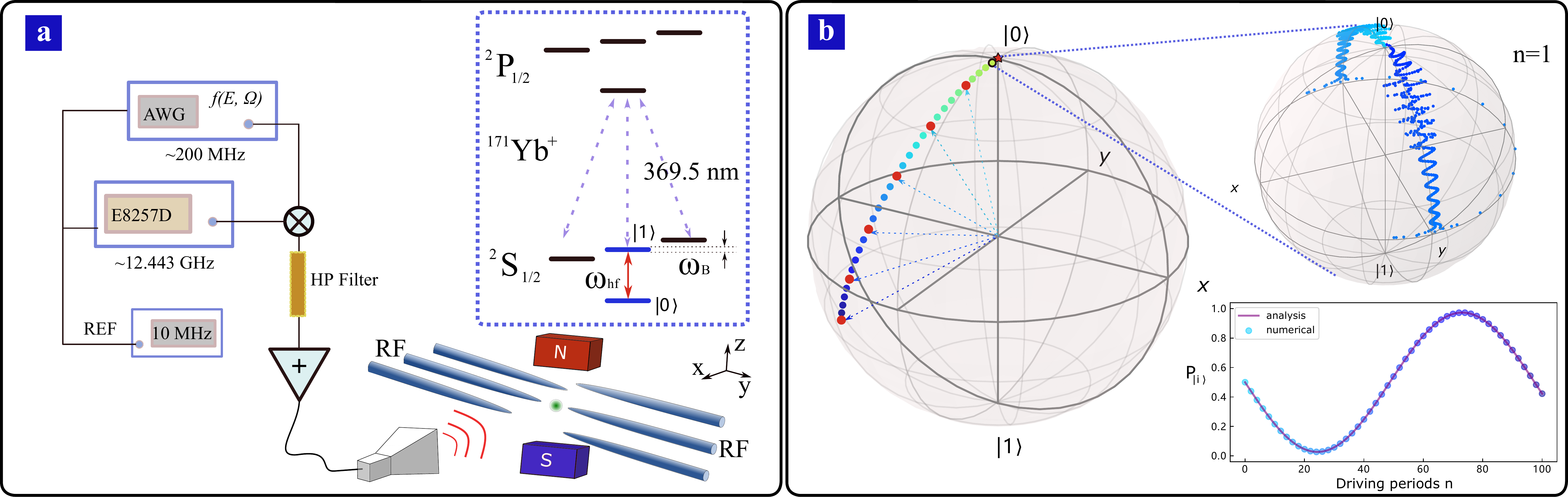}
\caption{\label{fig:setup_and_bloch} \textbf{Experimental procedure to measure
Floquet dynamics in a trapped-ion system}.  \textbf{a)} The qubit is encoded in
the clock transition of a single trapped $^{171}\mathrm{Yb}^{+}$ion. The qubit is periodically driven by a microwave
field generated by an AWG and a fixed-frequency signal source, which
are both referenced to a 10 MHz rubidium frequency standard.
\textbf{b)} On the right Bloch sphere, the qubit is initialized to $\left|0\right\rangle $ and evolves to $U(T)\left|0\right\rangle $ after one period of driving with function $f(E=16,\Omega=5)$. The fast dynamics (``micromotion'') of the qubit during this elementary
period is shown on the Bloch sphere, 
and arises from the intrinsic time-dependence of the Floquet states $\left|\Phi_{j}(t)\right\rangle$. On the left Bloch sphere, the qubit starts from $\left|0\right\rangle $ (the north pole) and 
evolves to $U(nT)\left|0\right\rangle$ for multiple periods $n=1, 2, 3, ..., 30. $ The red dots are the points measured at $n=5, 10, 15, 20, 25, 30$. The insert shows 
the probability $P_{\left|i\right\rangle }(n)$ when the qubit is measured onto base $\left|i\right\rangle =\frac{1}{\sqrt{2}}(\left|0\right\rangle +i\left|1\right\rangle )$ after n periods.
$P_{\left|i\right\rangle }(n)$ is a sinusoidal function
of $n$, with frequency proportional to the quasienergy $\epsilon_{j}(E)$.
For values of $E$ satisfying $\epsilon_{j}(E)=0$, the state is frozen
at $\left|0\right\rangle $ and $P_{\left|i\right\rangle }(n)=1/2$
for all $n$. Since $\epsilon_{j}(E)$ is proportional to the real part of
$-\zeta(1/2 + i E)/(1/2 + i E)$, we can therefore identify Riemann zeros by observing
the freezing of the dynamics, i.e. $P_{\left|i\right\rangle }=1/2$.
 }
\end{figure*}

\subsection{Experiment}
The experimental results were obtained by periodically driving
a single trapped ion with microwave fields. The two-level
system is encoded in the hyper-fine clock transition $\left|0\right\rangle \equiv{}^{2}S_{1/2}\left|F=0,\thinspace m_{F}=0\right\rangle $
and $\left|1\right\rangle \equiv{}^{2}S_{1/2}\left|F=1,\thinspace m_{F}=0\right\rangle $
in a single ytterbium ($^{171}\mathrm{Yb}^{+}$) ion confined in a
Paul ion trap \cite{cui2016experimental}, as shown
in Fig. \ref{fig:setup_and_bloch}\textbf{a}. This clock qubit has
the advantages of high-fidelity quantum operations and long coherence
time \cite{OM07, WK17}. 
The tunneling, $J$, in this system is of the order of
$10$kHz, giving a resonant Rabi time of $\sim 100 \mu$s. The driving function is switched on by fast modulating the detuning frequency.

After 1 ms of Doppler cooling and 50 $\mu$s of optical pumping,
the ion is initialized in the ground state $\left|0\right\rangle $
with a probability $\geq 99.5\%$. The qubit is then driven by a microwave
field for multiple periods. The driving function was generated from
a programmable arbitrary waveform generator (AWG) by phase modulating
a 200 MHz microwave sinusoidal signal with the driving function $f(t)$.
It is then mixed with a 12.4 GHz fixed frequency signal. The amplified
microwave fields were delivered to the trapped ion from an horn antenna
located outside the vacuum chamber. At the end of the multiple periods,
the state is measured in the basis $\left|i\right\rangle =\frac{1}{\sqrt{2}}(\left|0\right\rangle +i\left|1\right\rangle )$
by applying a $\pi/4$ rotation and normal fluorescence detection.
When more than one photon is detected, the measurement result is noted
as 1; otherwise it is noted as 0. The time evolution of the state
population is recorded as a function of the number of periods.

In Fig. \ref{fig:setup_and_bloch}\textbf{b} we show the experimental
protocol, plotted on the Bloch sphere.
As noted previously, the Floquet modes 
$\left|\Phi_{1}(E,\Omega,t)\right\rangle =a(t)\left|0\right\rangle +b(t)\left|1\right\rangle, 
\left|\Phi_{2}(E,\Omega,t)\right\rangle =b^*(t)\left|0\right\rangle - a^*(t)\left|1\right\rangle$  
are the eigenstates of the one-period time evolution operator
$U(E,\Omega)$, where $\Omega$ is the driving frequency, and $E$ is a driving
parameter related to the argument of the zeta function, $\zeta(1/2 + i E)$. Starting
from the initial state $\left|0\right\rangle $, the population of the state measured
in basis state $\left|i\right\rangle $ after $n$ periods of driving is
$P(n,E,\Omega)=1/2-A\sin(2nT\epsilon_{j}(E,\Omega))$, where $A=\mathrm{Re} \{a(0)b^*(0)\}$,
and $\epsilon_{j}(E,\Omega)$
is the quasienergy.
It is clear that if $E$ is equal to the zeros of  $\epsilon_{j}(E,\Omega)$, $\epsilon_{j}(E,\Omega)$ vanishes
and $P(n,E,\Omega)=1/2$ for all $n$. While if $\epsilon_{j}(E,\Omega)\neq0$,
$P(n,\Omega,E)$ evolves sinusoidally with a frequency proportional
to the quasienergy $\epsilon_{j}(E,\Omega)$. The the Riemann zeros
can thus be identified by observing the freezing of the evolution of $P(n,E,\Omega)$ produced by CDT.

\begin{figure*}
\includegraphics[width=1\linewidth]{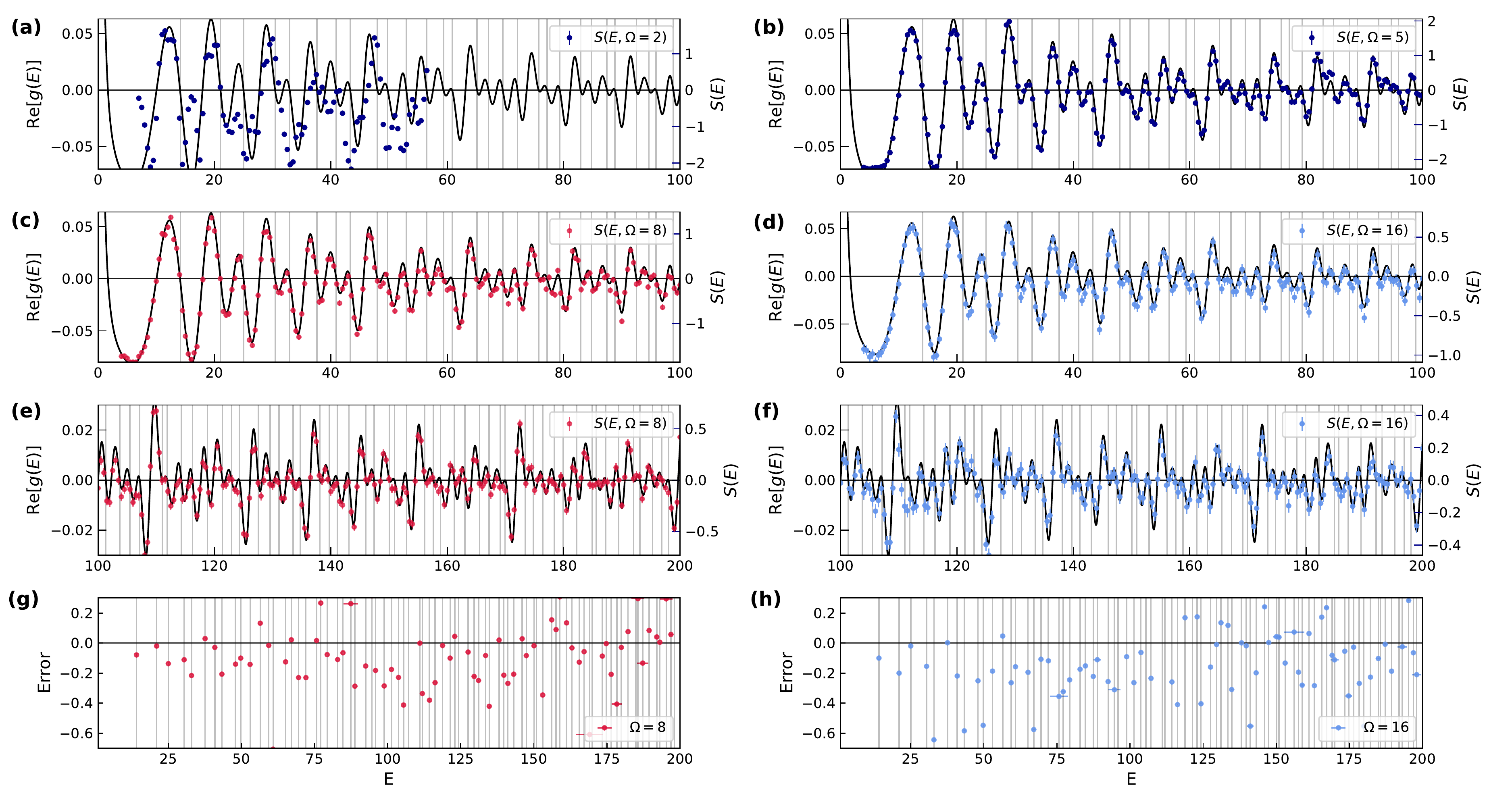}

\caption{\label{fig:Results}\textbf{Identifying Riemann zeros by observing
the frozen dynamics of the state.} In \textbf{(a-f)}, the black curve is the real part of  $g(E)$. The vertical grey lines indicate the Riemann zeros, \{14.1347, 21.022, 25.0109, 30.4249, 32.9351, ...\}; 
the other points where $g(E)$ crosses the axis
correspond to the real part of the zeta function vanishing but not the imaginary part, and so do not represent Riemann zeros. The dots are the $S$ parameter which is defined as $S(\Omega,E)=\sum_{n}[P_{\left|i\right\rangle }(n,\Omega,E)-1/2]$,
where $n=\{5,10,15,20,25,30\}$. $S(\Omega,E)$ is used to identify the Riemann zeros
by observing $S=0$.  \textbf{(a)} When $\Omega = 2$, the high-frequency limit is not well satisfied, so the behaviour of $S(\Omega,E)$ does not match $g(E)$.  \textbf{(b)} By contrast, $\Omega = 5$ does give good agreement for $E \leq 100$. \textbf{(c)} Increasing $\Omega$ further to 8 provides even better
agreement with $g(E)$. \textbf{(e)} We implemented the measurement for $E$ up to 200 and identified the first 80 Riemann zeros with high accuracy. \textbf{(d, f)} Results of $\Omega = 16$. The measurement can be extended to higher $E$ without loss of efficiency and accuracy.  Data points with $E \leq 100$ ($E \geq 100$) were obtained by 2000 (5000) measurements. The error bar $\delta S$ of  $S(\Omega,E)$ is the sum of the statistical errors of the corresponding $P_{\left|i\right\rangle }(n,\Omega,E)$, where $n=\{5,10,15,20,25,30\}$, within one standard deviation. 
\textbf{(g, h)} Zeros were extracted by interpolating $S(E,\Omega)$ using a cubic polynomial 4000 times. Each time, $S$ is sampled randomly in [$S-\delta S$, $S+\delta S$]. The extracted zeros are the mean values of the interpolated zeros (see Supplementary Material). There are a few vertical lines with no red points on them. This means that the error is so large that the points don't fall inside the plot range. The error (dot) is the difference between the extracted zeros and the exact zeros, with the error bars indicating the standard deviations of the mean values in the interpolation. 
}
\end{figure*}

To give a quantitative characterization of the state evolution, we define the $S$ parameter as $S(\Omega,E)=\sum_{n}[P_{\left|i\right\rangle }(n,\Omega,E)-1/2]$,
where $n=\{5,10,15,20,25,30\}$ are the number of driving periods
in the experiment. It is straightforward to see that the zeros of the
quasienergies are the zeros of $S(\Omega,E)$ as well. Therefore
a scan of the parameter $E$ allows us to identify the Riemann zeros
by observing $S(\Omega,E)=0$. To give a direct comparison,
we show in Fig. \ref{fig:Results} the experimental results of $S(\Omega,E)$ and the real part of $g(E)$ as a function of $E$ for $\Omega=2$, 5, 8, and 16, respectively. For $\Omega=2$, we can see that $S(\Omega,E)$ shows a significant distortion
from the theoretical zeta function and does not
allow the position of the Riemann zeros to be determined.
Increasing $\Omega$, however, substantially improves the results, and
gives excellent agreement
between data and theory over the range $E \leq 200$, allowing the first
eighty Riemann zeros to be resolved.
This improvement occurs because larger values of $\Omega$ satisfy the high-frequency approximation
better, and thus Eq. \ref{effective} becomes more precise. The difference between the measured zeros and the exact Riemann zeros is shown in Fig. \ref{fig:Results} (\textbf{g}). Most of the zeros can be identified with an accuracy of 1 \% or better. In Table \ref{values} we present the agreement quantitatively for four different driving frequencies over a wide range of $E$. 
We give further, and more detailed, comparisons of the agreement
in the Supplementary Material.

\begin{table}
\begin{center}
\begin{tabular}{ |l||c|c|c|c|c| }
 \hline
{ } & $E_1$ & $E_{10}$ & $E_{30}$ & $E_{50}$ & $E_{70}$\\
\hline
Exact & 14.135 & 49.774 & 101.318 & 143.112 & 182.207  \\
\hline
$\Omega = 5$ & 14.07(1) & 49.26(19) &  &  &  \\
$\Omega = 8$ & 14.06(2) & 49.67(3) & 101.13(3) &142.90(9) & 182.28(6)  \\
$\Omega = 12$ & 13.99(4) & 49.36(23) & 101.31(3) & 142.72(22) & 182.14(6)  \\
$\Omega = 16$ & 14.03(3) & 49.23(22) & 101.33(5) & 142.91(13) & 181.98(8)  \\
 \hline
\end{tabular}
\caption{{\bf Comparison of the experimentally measured Riemann zeros with the true values
for different driving frequencies.} Zeros were extracted by interpolating
$S(E,\Omega)\pm\delta S$ with a cubic polynomial 4000 times, where $\delta S$ is the $1-\sigma$ standard deviation of $S$. Each time, $S$ is sampled randomly in [$S-\delta S$, $S+\delta S$]. The zeros are the mean of the interpolated zeros. The values in parentheses denote the standard deviation of the means in terms of the least significant digit. 
}
\label{values}
\end{center}
\end{table}
Since increasing $\Omega$ further would satisfy the high-frequency limit better, it might be thought that the accuracy of the results can be improved by increasing
the driving frequency to arbitrarily high values.
This is not the case however.
As we show in {\bf Methods}, the best results will be obtained when the driving period $T$ is small
enough to satisfy the high-frequency limit, while at the same time $T$ is sufficiently large
for it to replace the upper limit of integration in Eq. \ref{final}. As a consequence of
these opposing requirements, the best results will actually be obtained for mid-range frequencies.
In Fig. \ref{fig:Results} (\textbf{d}, \textbf{f} and \textbf{h}) we show the results for $\Omega = 16$. Comparing with the $\Omega = 8$ result reveals that increasing
the frequency has not improved the accuracy of the results.


\subsection{Reconstruction of prime numbers}

\begin{figure}
\includegraphics[width=0.47\linewidth]{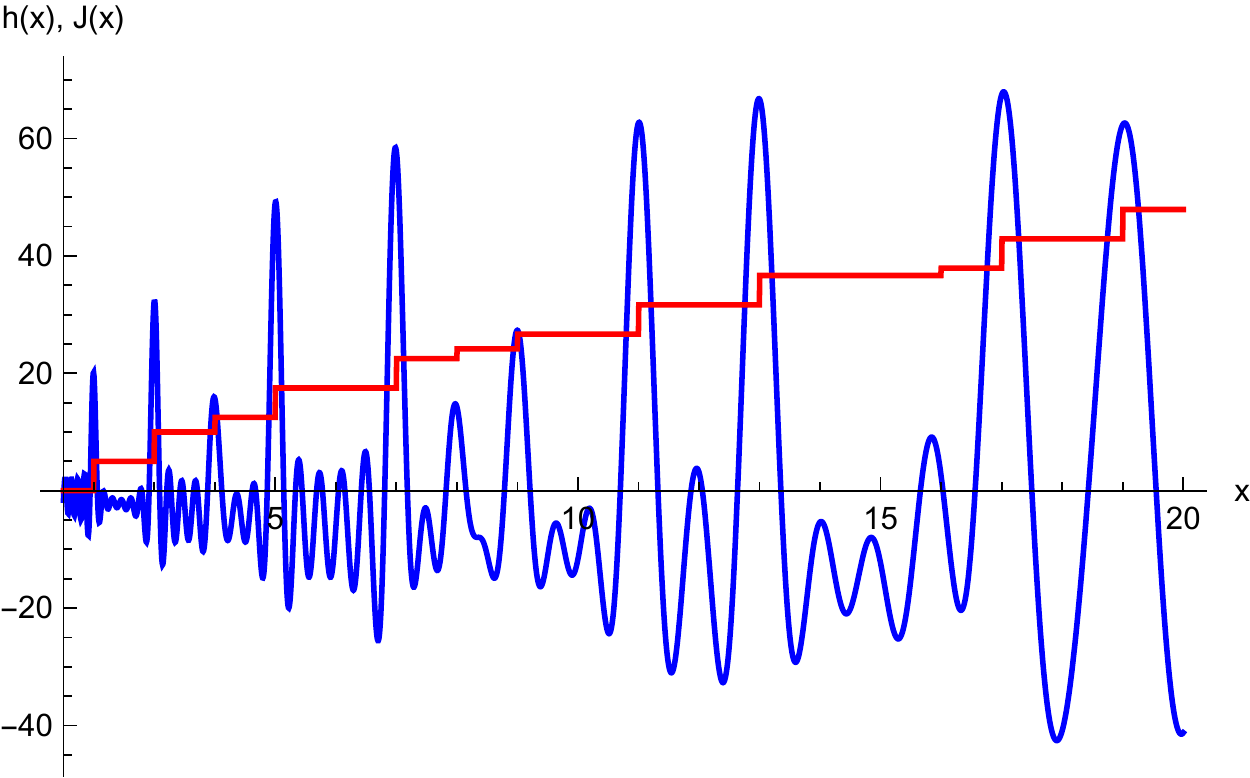}
\includegraphics[width=0.47\linewidth]{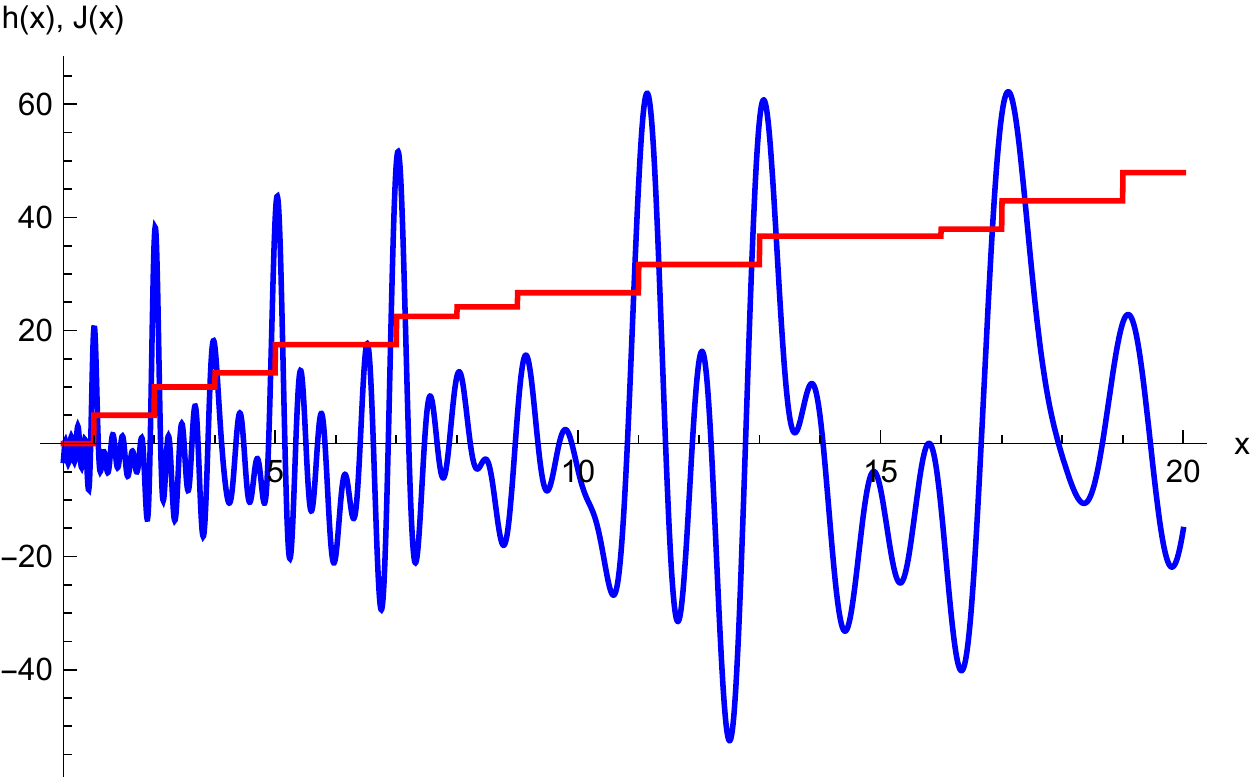}
\caption{\label{fig:hJ}{\bf Primes from zeros.} Plot of the function Eq. \eqref{hx} 
with  the sum restricted to $|E_n| < 100$ (blue) and the function $5 J(x)$ given in  Eq. \eqref{Jx} (red).
(a) using the exact values of $E_n$. (b) using  the values of $E_n$ given in Table II (Extended Data) for $\Omega =16$.  
In both cases one can identify the first eight primes and their powers. 
}
\end{figure}

In 1859 Riemann found a formula that gives  the number of primes $\pi(x)$ below or equal to $x$ in terms of the 
non trivial zeros $\rho_n = \frac{1}{2} + i E_n$  \cite{riemann}. A consequence of this result is that the function  \cite{conrey} 
\begin{equation}
h(x) = - \sum_{\rho} x^{\rho} 
\label{hx}
\end{equation}
has peaks at the primes $p$ and their powers $p^n$. 
Fig. \ref{fig:hJ} shows a truncation of \eqref{hx}  together with the function
\begin{equation} 
J(x) = \sum_{n \geq 1} \frac{1}{n} \pi(x^{1/n})
\label{Jx}
\end{equation}
that jumps by 1 at every prime and by $\frac{1}{n}$ at the power $p^n$. 
Notice that the experimental error in the {\em zeros}  does not affect 
appreciably the location of the peaks.

\subsection{Conclusions}
We have presented an experimental method for measuring the location of the
zeros of the Riemann $\zeta$-function, by using Floquet engineering to control the
quasienergy levels of a periodically-driven trapped ion. 
The experimentally measured
values of the zeros are in excellent agreement with their theoretical values,
and we have demonstrated how they can be used to reconstruct the
prime numbers.
The high level of experimental control over this system, and the implementation
of a driving function derived from the complex function
$g(E)= - \zeta(z)/z$ (where $z = 1/2 + i E$), 
allows as many as the first 80 {\em zeros}  to be resolved.
Our analysis indicates that there is a ``sweet spot'' for the driving frequency,
in which $\Omega$ is sufficiently large for the system to be in the
high-frequency regime, while its period is large in comparison to the
width of the Fourier transform of $g(z)$. 
Using the experimentally measured {\em zeros} we have also obtained a
good approximation of the lowest prime numbers. This reconstruction
suggests the possibility of a direct experimental realization of the primes.
The successful realization of the Riemann zeros in a quantum mechanical system
represents an important step 
along a route inspired by the Hilbert and P\'olya proposal,
and may lead to further insights into the Riemann hypothesis.

\begin{figure*}
\includegraphics[width=1\linewidth]{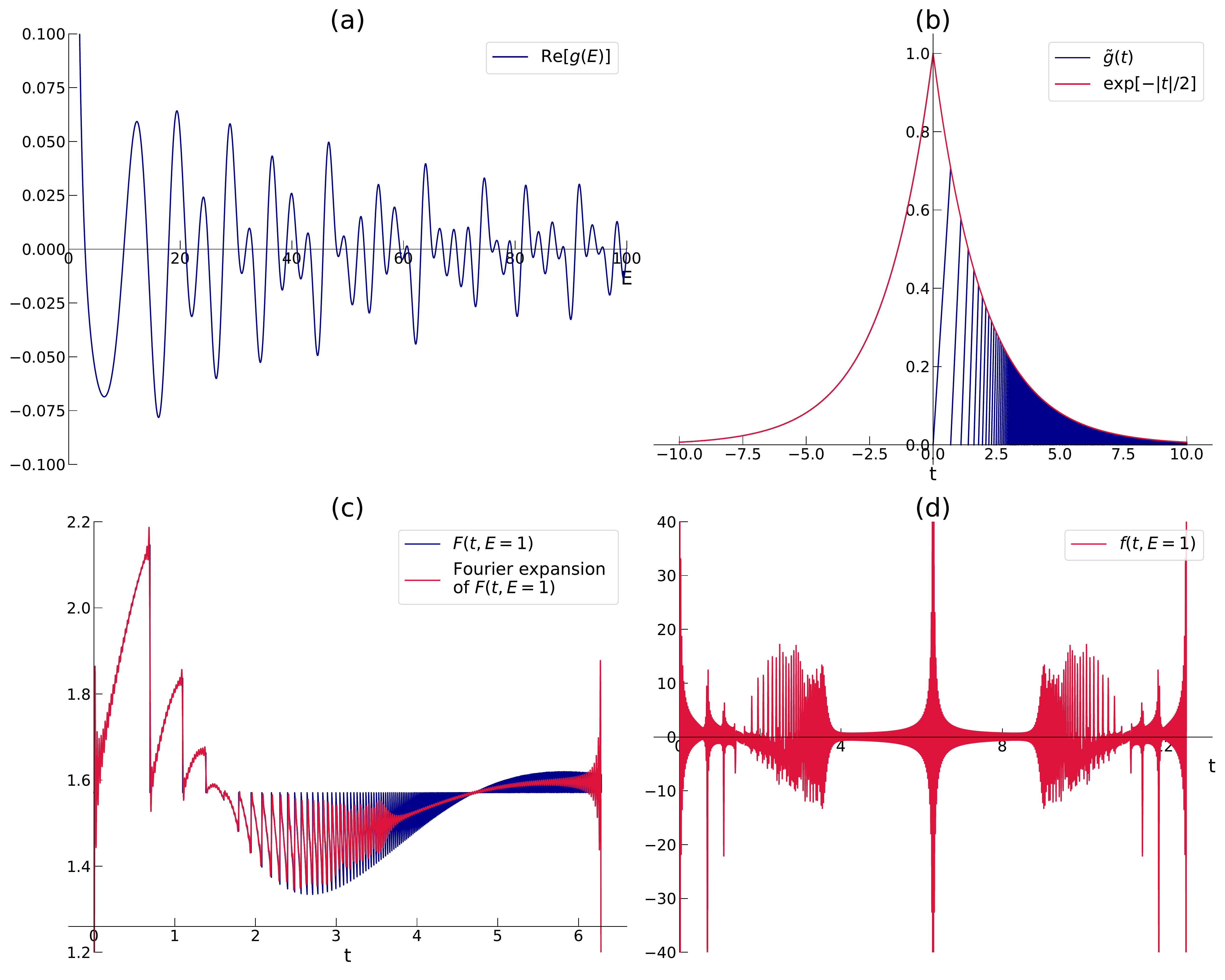}

\caption{\label{fig:math}{\bf Derivation of the driving function.} (a) The real component of $g(E)= - \zeta(1/2 + i E)/(1/2 + i E)$.
(b) Van der Pol's function is the Fourier transform, 
${\tilde g}(t)$, of $g(E)$. The function is
bounded by the red curve $\mathrm{exp} \left[ - |t| / 2 \right]$, and contains an infinite
number of finite discontinuities for positive $t$, arising from the floor function
(see Eq. \ref{van_der_pol}).
(c) The primitive of the driving function, $F(t)$, for driving parameter $E = 1$.
The discontinuities in ${\tilde g}(t)$ give rise to discontinuities in this
function as well. The red curve shows the Fourier expansion of $F(t)$, truncated at
500 terms. We can note how the fine detail is progressively blurred out as $t$ increases.
(d) The driving function, $f(t)$, for $E=1$, obtained as $f(t) = \partial_t F(t)$,
plotted over $0 \leq t < T$.
The discontinuities in $F(t)$ produce $\delta$-function spikes in the driving function.
By construction, $f(t)$ is an even function of $t$, and so the full periodicity of this
function is $2 T$.
}
\end{figure*}

\bigskip
\section{Methods}

\subsection{Driving function derivation}
Our starting point is
the function $g(E) = \frac{ - \zeta\left( \mbox{\nicefrac{1}{2}} + i E \right)}
{\mbox{\nicefrac{1}{2}} + i E }$, where $\zeta(s)$ is the standard Riemann
zeta function. 
We plot the behaviour of this function in Fig. \ref{fig:math}a.
As van der Pol
showed in 1947, its Fourier transform  (Fig. \ref{fig:math}b)
can be written in the surprisingly simple form \cite{vdp} 
\begin{equation}
{\tilde g}(t) =  e^{t/2} - e^{- t/2}  [e^t]  \ .
\label{van_der_pol}
\end{equation}
where $[x]$ is the integer part of $x$. 
As we can see from Fig. \ref{fig:math}b, this function is localized around the origin,
with an envelope of the form $\mathrm{exp} \left(- | t | /2 \right)$.

By dividing the range of integration for the Fourier transform into two halves, it is
straightforward to show that the real component of $g(E)$ is given by
\begin{equation}
\mathrm{Re} \left[ g(E) \right] =
\frac{2}{4 E^2 + 1} + \int_{0}^{\infty} {\tilde g}(t) \cos E t \ dt \ .
\label{final}
\end{equation}
In order to observe the location of the Riemann zeros, our interest is
focused on values of $E > 10$. Accordingly we can simply discard the first term, as over this range
its magnitude is smaller than the experimental uncertainly in the measurements.

Our aim is to obtain a driving function $f(t)$ such that the effective tunneling is proportional
to the real component of $g(E)$, that is, $\mathrm{Re} \left[ J_\mathrm{eff} \right]
= \alpha \ \mathrm{Re} \left[g(E) \right]$, where $\alpha$ is the constant of proportionality. 
Comparing Eq.\eqref{final} with Eq.\eqref{effective}, and assuming that the driving period $T$
is sufficiently large to replace the upper limit of integration in \eqref{final},
reveals that $F(t) = \cos^{-1} \left( \alpha T {\tilde g}(t) \cos E t \right)$. 
The boundary condition $F(0) = 0$ requires
setting $\alpha = 1 / T$, which yields the final  
driving function $f(t) = \partial_t \left[ \cos^{-1} \left( {\tilde g}(t) \cos E t \right) \right]$. 
This choice of $\alpha$ also imposes the condition that the argument of the inverse 
cosine function is bounded within $\pm 1$ as required, since $ {\tilde g}(0)$ is the 
global maximum of $ {\tilde g}(t)$.
We can note that replacing the upper limit of integration with $T$ represents an important restriction
on the value of $\Omega$. 
This replacement means that $T$ must be large in comparison with the width of
${\tilde g}(t)$, and thus the driving frequency $\Omega$ must correspondingly be low.
However, for Eq.\eqref{effective} to be an accurate description of the system's dynamics requires
a high value of $\Omega$, so that the system is in the high-frequency regime.
Therefore, good results will be obtained in an
intermediate range of frequency, when both of these conditions can
be adequately satisfied.

We show the form of the $F(t)$ and the driving function for a particular value of $E$ in
Fig. \ref{fig:math}c and Fig. \ref{fig:math}d. The finite discontinuities present in $g(E)$
also produce discontinuities in $F(t)$, and thus $\delta$-function spikes
in $f(t)$. A convenient way to obtain $f(t)$ numerically is to expand
$F(t)$ in a Fourier series, differentiate the series term by term, and then to re-sum it. 
As in Ref.\cite{pra_1}, we want the driving function to be of definite parity, 
so that the two Floquet states will be of opposite parity, and so can
cross as the driving parameter $E$ is varied. If this parity condition were not satisfied, the 
von Neumann-Wigner theorem would prevent the quasienergies becoming degenerate, and they
could only form broader avoided crossings instead.
For this reason we choose to expand $F(t)$ as a Fourier sine series, so that
its derivative, $f(t)$ is a cosine series, and is thus an even function of time.
Sufficient terms must be included in the series to ensure that the fine structure in $f(t)$
is reproduced with sufficient resolution. Typically in the experiment the series was truncated at 500
terms.

\subsection{Experimental details}
A long coherence time of the system is vital in the
experiment. The hyperfine splitting of the  ($^{171}\mathrm{Yb}^{+}$) ion, 
$\omega_{hf}=\left( 12642812118.5+\omega_{B} \right)$
Hz, has a second-order Zeeman shift $\omega_{B}=310.8B^{2}$,
where $B$ is the magnetic field. 
We used $\mathrm{Sm}_{2}\mathrm{Co}_{17}$ permanent magnets to generate a static magnetic field
of around $B=9.15$ G to reduce the 50 Hz ac-line noise. The whole platform is shielded in a 2-mm-thick $\mu$-metal enclosure to reduce the
residual fluctuating magnetic fields \cite{RP16}. During the experiment, we still observed a slow drift of $\sim\pm30$
Hz of the clock transition in 10 hours. This
corresponds to $\Delta B\sim\pm0.005$ G, which is mainly due to the temperature drift in the laboratory. This drift is not negligible. Therefore the clock transition
frequency was frequently measured by Ramsey type measurements and calibrated
by updating the AWG wave frequency during the experiment every half hour.

\section{Data availability}
Source data and all other data that support the plots within this paper and other findings of this study are available from the corresponding author upon reasonable request.

{\bf Acknowledgments}
CEC was supported by the Spanish MINECO through grant FIS2017-84368-P,
and GS by PGC2018-095862-B-C21, QUITEMAD+ S2013/ICE-2801, SEV-2016-0597 and the CSIC Research Platform on Quantum Technologies PTI-001. RH, MZA, JMC, YFH, YJH, CFL, and GCG were supported
by the National Key Research and Development Program
of China (Grants No.2017YFA0304100 and No. 2016YFA0302700), the National Natural Science Foundation
of China (Grants No.11874343, No. 61327901,
No. 11774335, and No. 11734015), Key Research Program of Frontier Sciences,
CAS (Grant No. QYZDY-SSW-SLH003), the Fundamental
Research Funds for the Central Universities
(Grants No. WK2470000026, No. WK2470000027, and
No. WK2470000028), and Anhui Initiative in Quantum
Information Technologies (Grants No. AHY020100 and
No. AHY070000).

{\bf Author Information}
CEC and GS developed the theoretical proposal. RH, MZA, JMC and YFH designed and performed the experiment. YJH, CFL and GCG supervised the experiments. All authors
contributed to the data analysis, progression of the project, discussion of the results and the writing of the manuscript.
\pagebreak{}

\clearpage{}



\begin{figure*}

\section*{Extended Data}

\includegraphics[width=0.8\linewidth]{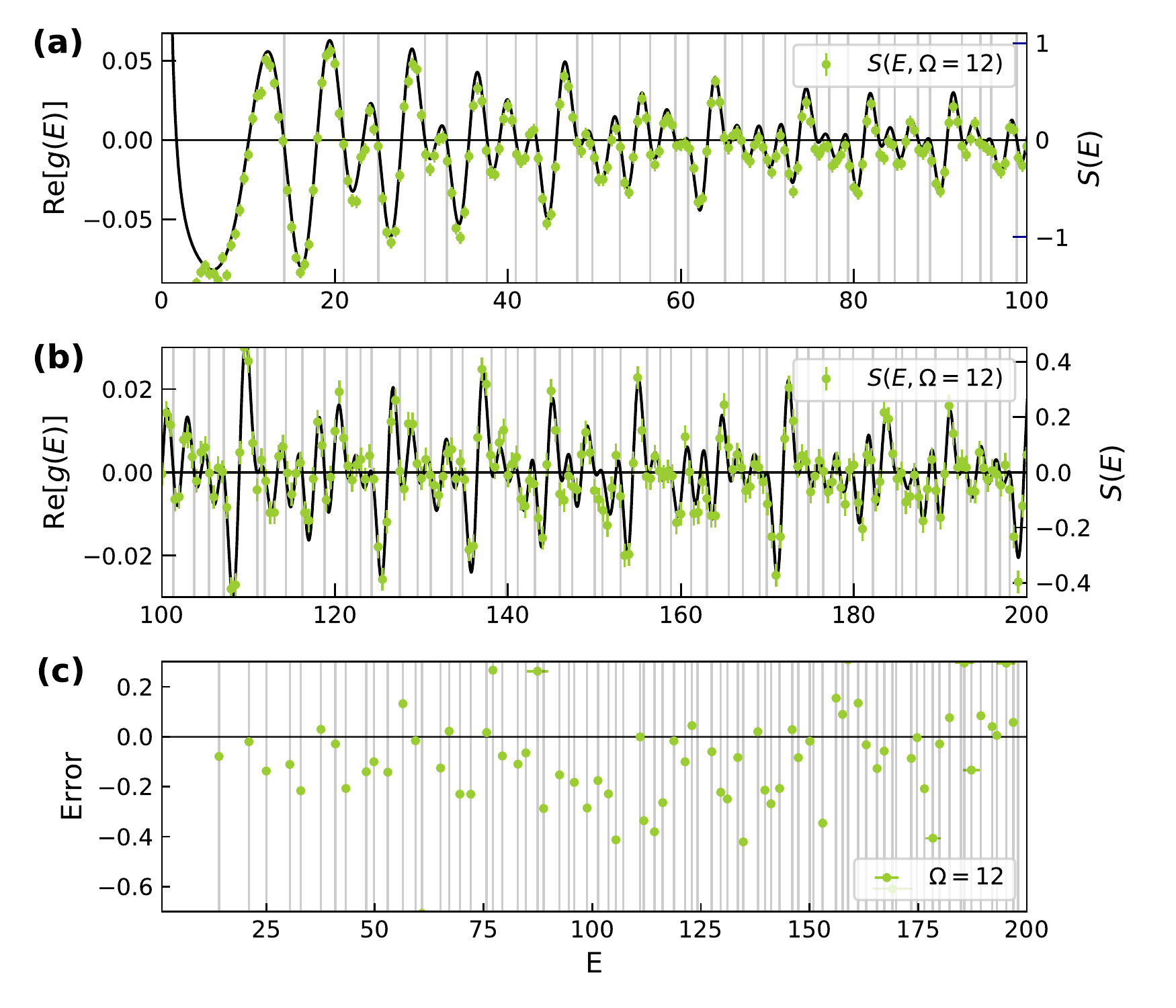}

\caption{\label{fig:Results12}\textbf{Identifying Riemann zeros by observing
the frozen dynamics of the state for $\Omega = 12$.} In \textbf{(a, b)}, the black curve is the real part of  $g(E)$. The vertical grey lines indicate the Riemann zeros, e.g., \{14.1347, 21.022, 25.0109, 30.4249, 32.9351, ...\}. 
The other points where $g(E)$ crosses the zero axis
correspond to the real part of the zeta function vanishing but not the imaginary part, and so do not represent Riemann zeros. The dots are the $S$ parameter which is defined as $S(\Omega,E)=\sum_{n}[P_{\left|i\right\rangle }(n,\Omega,E)-1/2]$,
where $n=\{5,10,15,20,25,30\}$. $S(\Omega,E)$ is used to identify the Riemann zeros
by observing $S=0$.  
$S(\Omega = 12, E)$ shows an excellent consistency in behavior with $g(E)$ for both \textbf{(a)} $0 \leq E \leq 100$ and \textbf{(b)} $100 \leq E \leq 200)$ and thus acts as a good indicator of the location of the Riemann zeros . The first 80 Riemann zeros can be identified with high accuracy. The measurement can be extended to higher $E$ without loss of efficiency and accuracy.  Data points with $E \leq 100$ ($E \geq 100$) were obtained by 2000 (5000) measurements. The error bar $\delta S$ of  $S(\Omega,E)$ is the sum of the statistical errors of the corresponding $P_{\left|i\right\rangle }(n,\Omega,E)$, where $n=\{5,10,15,20,25,30\}$, within one standard deviation. 
\textbf{(c)} Zeros were extracted by interpolating $S(E,\Omega)$ using a cubic polynomial 4000 times. Each time, $S$ is sampled randomly in [$S-\delta S$, $S+\delta S$]. The extracted zeros are the mean values of the interpolated zeros (see Table S1 and Table S2). The error (dot) is the difference between the extracted zeros and the exact zeros, with the error bars indicating the standard deviations of the mean values in the interpolation.
}
\end{figure*}

\begin{table*}

\begin{center}
\begin{tabular}{ |l||c|c|c|c|c| }
\hline
No.& Exact & $\Omega=5$ & $\Omega=8$ & $\Omega=12$ & $\Omega=16$\\
\hline
1 & 14.135 & 14.07(1) & 14.06(2) & 13.99(4) & 14.03(3) \\
2 & 21.022 & 21.04(2) & 21.00(2) & 20.93(5) & 20.82(3) \\
3 & 25.011 & 24.70(3) & 24.87(2) & 24.87(7) & 24.99(4) \\
4 & 30.425 & 30.59(2) & 30.31(2) & 30.29(3) & 30.27(4) \\
5 & 32.935 & 32.76(3) & 32.72(3) & 32.57(8) & 32.29(23) \\
6 & 37.586 & 37.64(2) & 37.62(2) & 37.39(2) & 37.59(4) \\
7 & 40.919 & 40.95(2) & 40.89(3) & 40.78(3) & 40.70(4) \\
8 & 43.327 & 42.85(9) & 43.12(4) & 43.23(4) & 42.74(40) \\
9 & 48.005 & 48.23(4) & 47.87(6) & 47.94(6) & 47.75(9) \\
10 & 49.774 & 49.26(19) & 49.67(3) & 49.36(23) & 49.23(22) \\
11 & 52.970 & 52.93(2) & 52.83(4) & 52.88(5) & 52.78(5) \\
12 & 56.446 & 56.56(3) & 56.58(3) & 56.28(3) & 56.49(5) \\
13 & 59.347 & 59.44(5) & 59.33(9) & 59.35(6) & 59.08(28) \\
14 & 60.832 & 60.10(48) & 60.13(414) & 60.41(144) & 60.67(9) \\
15 & 65.113 & 65.53(11) & 64.99(4) & 65.05(6) & 64.92(6) \\
16 & 67.080 & 67.06(5) & 67.10(3) & 66.98(10) & 66.50(40) \\
17 & 69.546 & 69.36(4) & 69.32(7) & 69.11(28) & 69.44(7) \\
18 & 72.067 & 71.82(3) & 71.84(3) & 71.76(8) & 71.95(7) \\
19 & 75.705 & 76.33(37) & 75.72(12) & 75.23(332) & 75.35(317) \\
20 & 77.145 & 76.84(8) & 77.41(6) & 76.80(9) & 76.82(83) \\
21 & 79.337 & 78.89(6) & 79.26(2) & 78.95(4) & 79.09(9) \\
22 & 82.914 & 84.12(117) & 82.80(2) & 82.67(4) & 82.74(8) \\
23 & 84.736 & 84.82(2) & 84.67(3) & 84.31(12) & 84.58(8) \\
24 & 87.425 & 87.50(7) & 87.69(244) & 87.23(6) & 87.20(8) \\
25 & 88.809 & 87.93(47) & 88.52(3) & 88.33(47) & 88.70(128) \\
26 & 92.492 & 92.96(4) & 92.34(3) & 92.37(5) & 92.24(6) \\
27 & 94.651 & 94.55(59) & 94.97(12) & 94.34(144) & 94.34(211) \\
28 & 95.871 & 95.82(8) & 95.69(5) & 94.66(290) & 95.13(125) \\
29 & 98.831 & 98.87(3) & 98.55(6) & 98.69(4) & 98.74(6) \\
 \hline
\end{tabular}
\caption{{\bf Comparison of the experimentally measured Riemann zeros with the true values ($1\leq E\leq100$) 
for different driving frequencies.} Zeros were extracted by interpolating
$S(E,\Omega)\pm\delta S$ with a cubic polynomial 4000 times, where $\delta S$ is the $1-\sigma$ standard deviation of $S$. Each time, $S$ is sampled randomly in [$S-\delta S$, $S+\delta S$]. The zeros are the mean of the interpolated zeros. The values in parentheses denote the standard deviation of the means in terms of the least significant digit. 
}
\label{values100}
\end{center}
\end{table*}

\begin{table*}
\begin{center}
\begin{tabular}{ |l||c|c|c|c| }
\hline
No.& Exact & $\Omega=8$ & $\Omega=12$ & $\Omega=16$\\
\hline
30 & 101.318 & 101.13(3) & 101.31(3) & 101.33(5) \\
31 & 103.726 & 103.50(4) & 103.74(9) & 103.66(5) \\
32 & 105.447 & 105.03(16) & 105.49(7) & 104.46(58) \\
33 & 107.169 & 106.03(112) & 106.99(17) & 106.93(5) \\
34 & 111.030 & 111.03(16) & 111.60(227) & 110.27(173) \\
35 & 111.875 & 111.54(19) & 111.88(7) & 112.86(337) \\
36 & 114.320 & 113.94(10) & 114.49(9) & 114.06(6) \\
37 & 116.227 & 115.96(11) & 116.12(6) & 115.82(20) \\
38 & 118.791 & 118.77(3) & 118.71(4) & 118.96(7) \\
39 & 121.370 & 121.27(6) & 121.78(97) & 122.26(69) \\
40 & 122.947 & 122.99(22) & 123.48(323) & 123.12(5) \\
41 & 124.257 & 122.82(48) & 124.43(19) & 123.85(38) \\
42 & 127.517 & 127.46(3) & 127.51(4) & 127.36(7) \\
43 & 129.579 & 129.36(32) & 129.66(12) & 129.57(12) \\
44 & 131.088 & 130.84(16) & 130.91(156) & 131.22(5) \\
45 & 133.498 & 133.41(7) & 134.00(36) & 133.62(12) \\
46 & 134.757 & 134.34(24) & 134.89(132) & 134.45(27) \\
47 & 138.116 & 138.14(8) & 138.27(10) & 138.12(11) \\
48 & 139.736 & 139.52(6) & 139.98(9) & 139.72(9) \\
49 & 141.124 & 140.86(8) & 141.20(5) & 140.57(107) \\
50 & 143.112 & 142.90(9) & 142.72(22) & 142.91(13) \\
51 & 146.001 & 146.03(8) & 145.80(4) & 146.24(52) \\
52 & 147.423 & 147.34(8) & 147.17(46) & 147.43(11) \\
53 & 150.054 & 150.04(6) & 149.62(73) & 150.10(125) \\
54 & 150.925 & 150.20(414) & 149.47(197) & 150.96(22) \\
55 & 153.025 & 152.68(6) & 152.79(5) & 152.89(14) \\
56 & 156.113 & 156.27(10) & 156.02(203) & 156.19(343) \\
57 & 157.598 & 157.69(20) & 157.60(89) & 157.40(15) \\
58 & 158.850 & 159.16(379) & 158.84(89) & 158.57(79) \\
59 & 161.189 & 161.32(8) & 161.00(4) & 161.25(3) \\
60 & 163.031 & 163.00(5) & 162.43(58) & 162.75(12) \\
61 & 165.537 & 165.41(5) & 165.94(26) & 165.71(14) \\
62 & 167.184 & 167.13(5) & 167.08(11) & 167.42(85) \\
63 & 169.095 & 168.49(458) & 169.16(49) & 169.01(14) \\
64 & 169.912 & 169.18(67) & 169.17(27) & 169.80(138) \\
65 & 173.412 & 173.33(5) & 173.60(19) & 173.36(8) \\
66 & 174.754 & 174.75(3) & 174.65(7) & 174.40(105) \\
67 & 176.441 & 176.23(11) & 176.52(16) & 176.41(13) \\
68 & 178.377 & 177.97(175) & 178.26(10) & 178.11(11) \\
69 & 179.916 & 179.89(5) & 180.01(262) & 179.36(51) \\
70 & 182.207 & 182.28(6) & 182.14(6) & 181.98(8) \\
71 & 184.874 & 185.51(294) & 184.82(17) & 184.77(86) \\
72 & 185.599 & 185.89(223) & 185.43(24) & 184.60(328) \\
73 & 187.229 & 187.10(188) & 187.04(72) & 187.22(34) \\
74 & 189.416 & 189.50(0) & 189.28(6) & 189.23(6) \\
75 & 192.027 & 192.07(9) & 192.20(169) & 192.42(41) \\
76 & 193.080 & 193.09(4) & 193.10(12) & 193.06(159) \\
77 & 195.265 & 195.56(217) & 195.18(80) & 195.55(58) \\
78 & 196.876 & 196.93(28) & 196.04(304) & 196.81(9) \\
79 & 198.015 & 196.81(42) & 197.74(10) & 197.80(154) \\
80 & 201.265 & 199.24(576) & 200.14(5) & 200.47(16) \\
 \hline
\end{tabular}
\caption{{\bf Comparison of the experimentally measured Riemann zeros with the true values ($100\leq E\leq200$) 
for different driving frequencies.} Zeros were extracted by interpolating
$S(E,\Omega)\pm\delta S$ with a cubic polynomial 4000 times, where $\delta S$ is the $1-\sigma$ standard deviation of $S$. Each time, $S$ is sampled randomly in [$S-\delta S$, $S+\delta S$]. The zeros are the mean of the interpolated zeros. The values in parentheses denote the standard deviation of the means in terms of the least significant digit. 
}
\label{values200}
\end{center}
\end{table*}

\end{document}